\documentclass[12pt,letterpaper]{article}
\usepackage[utf8]{inputenc}
\usepackage[english]{babel}
\usepackage{amsmath}
\usepackage{amsfonts}
\usepackage{amssymb}
\usepackage{graphicx} 
\usepackage{listings}
\usepackage{color}
\usepackage[left=2cm,right=2cm,top=2cm,bottom=2cm]{geometry}
\usepackage[tagged]{accessibility}
\usepackage[pdflang={en-US}]{hyperref}
\usepackage{xcolor}
\usepackage{siunitx}
\DeclareSIUnit\year{yr}

\allowdisplaybreaks

\newcommand{\authorstyle}[1]{{\large\usefont{OT1}{phv}{b}{n}\color{black}#1}} 

\newcommand{\institution}[1]{{\footnotesize\usefont{OT1}{phv}{m}{sl}\color{black}#1}} 

\usepackage{titling} 

\newcommand{\HorRule}{\color{gray}\rule{\linewidth}{1pt}} 

\pretitle{
	\vspace{-30pt} 
	\HorRule\vspace{10pt} 
	\fontsize{32}{36}\usefont{OT1}{phv}{b}{n}\selectfont 
	\color{black} 
}

\posttitle{\par\vskip 15pt} 

\preauthor{} 

\postauthor{ 
	\vspace{-10pt} 
	\par\HorRule 
	\vspace{20pt} 
}

\title{Ka-chow! A simple irregular 2D lattice model of lightning}

\author{
	\authorstyle{Gavin Buxton} 
	\newline\newline 
	\institution{Science Department, Robert Morris University, Moon Township, PA 15108, US.}\\ 
}

\date{}

\begin{document}

\maketitle 

\vspace{-80pt}
A model of lightning that captures the propagation of lightning channels on an irregular lattice is developed. The irregular lattice allows us to capture large two-dimensional systems (2 km $\times$ 2 km), while allowing grid refinement (on the order of cm) near areas of interest. Furthermore, the direction of lightning propagation is not biased in the orthogonal or diagonal directions of regular lattices. The probability of lightning strikes on people in standing positions, people in crouched positions, people near a tree and cattle are estimated. 

\section*{Introduction}

Lightning is the second deadliest weather phenomena, killing twenty-four thousand people each year \cite{ritenour2008lightning, christophides2017cardiac}.
However, this accounts for only 30\% of lightning victims, with 74\% of those that survive suffering from permanent disabilities \cite{cooper1980lightning, adekoya2005struck}; the rates of injuries are thought to be significantly under reported \cite{ritenour2008lightning}.
Injuries can consist of burns, cardiac effects and neurological consequences.
Ohmic heating as the high electrical current flows through the human body can cause burns, especially near metallic objects like jewelry or from clothing that might melt \cite{gookin2010backcountry, blumenthal2021injuries}.
In particular, the skin has a high electrical resistance \cite{maghsoudi2007electrical, arshad2018modeling} and so most current may travel over the skin (external flashover effect) \cite{ritenour2008lightning, blumenthal2021injuries}. 
This might result in the common Lichtenberg figures \cite{blumenthal2021injuries}, or it might not result in any obvious signs of being struck by lightning. That said, the large accompanying magnetic field variations can still cause secondary currents internally \cite{christophides2017cardiac}.
The most dangerous aspect of lightning is the ability of the large currents to cause cardiac arrhythmias, myocardium depolarisation, and cardiac arrest \cite{christophides2017cardiac, blumenthal2021injuries}. 
That said, patients are thought to often only require respiratory support and have been observed to respond well to cardiopulmonary resuscitation \cite{ritenour2008lightning}. 
The respiratory control center may take longer to recover from the lightning strike \cite{gookin2010backcountry}, and there may be other neurological effects such as numbness and paralysis \cite{gookin2010backcountry, blumenthal2021injuries}.
The large pressure changes during the lightning strike might also injure a person \cite{gookin2010backcountry}, especially ear barotrauma that can result in hearing loss \cite{blumenthal2021injuries}.
In lightning survivors there may even be ophthalmological (i.e., cataracts) and psychological (i.e., post-traumatic stress disorder) that do not present immediately \cite{blumenthal2021injuries}. \\

Lightning safety advice would always urge one to go indoors during a thunderstorm, but if that is not possible then it is advised to avoid contact with conducting structures, to not be in close proximity to large structures (e.g., trees) that may be hit, to seek lower ground, and to adopt the lightning crouch position \cite{roeder2008analysis}. That said, the vast majority of lightning fatalities in the US are thought to occur in close proximity to safe locations \cite{roeder2014backcountry}.
Although even going indoors may not provide 100\% protection against lightning injuries; the lightning is more likely to travel down electrical wiring and plumbing, but there is no certainty that that will be the case \cite{roeder2008analysis}.
When outdoors, one should avoid elevated terrain and be aware that lightning will most likely strike larger objects; objects twice as high are thought to be around 4 times as likely to be hit by lightning \cite{gookin2010backcountry}, although recent simulations have predicted a \qty{600}{\m} tall structure is 3.6 times as likely to be hit by lightning than a \qty{100}{\m} tall structure \cite{jiang2020simulation}. 
Conducting objects (e.g., metal tent poles or railings) may either channel the lightning strike or provide a path for ground currents and be the source of an upward leader \cite{gookin2010backcountry}.
Open areas where one might become the tallest object should also be avoided \cite{roeder2008analysis}.    
Once one is not too close to nearby trees, nor in a large clearing, and not on elevated ground they can reduce their chance of injury by adopting the lightning crouch position; squatting down low, with ones feet together and covering the ears \cite{roeder2008analysis, gookin2010backcountry}.
However, the relative risk of injury will depend on how the 30 to 50 kA of current is channeled near or through the body \cite{blumenthal2021injuries}.\\

Lightning injury due to a direct strike, when a positive upward leader initiated from a person makes contact with a downward negative leader, occurs in only a few percent of cases but might intuitively be expected to result in fatalities \cite{ritenour2008lightning, cooper2010distribution}.
20 to 30\% of lightning injuries occur due to side flashes, when lightning hits a nearby tall structure (e.g., a tree) and jumps to a nearby (more conducting) person \cite{gookin2010backcountry, cooper2010distribution}. Such side flashes may be influenced by the person holding a conducting object \cite{gomes2021metal}.
Lightning current may be channeled by conductors (e.g., telephone wires, plumbing, metal fencing, or hand rails) and 15 to 25\% of injuries occur because someone was in direct contact with the electrified object \cite{gookin2010backcountry, cooper2010distribution}.
The most dangerous scenario, accounting for 40 to 50\% of lightning fatalities \cite{cooper2010distribution}, involves a stepped potential. 
Lightning current flowing through high resistance soil near a lightning strike may flow in an alternative path; namely up one leg of the victim and down the other \cite{roeder2008analysis}.
This can be more fatal for animals if they are oriented with the voltage difference being greatest between the front and rear legs such that the current flows through their heart \cite{colton1950lightning, shaw1973electrocution, cristancho2017fatal}.
More recently a fifth mechanism of lightning injury, due to the development of a positive upward leader from the person, was proposed \cite{cooper2002fifth} that is thought to account for 10 to 15\% of lightning injuries \cite{cooper2010distribution}. 
Associated effects of lightning strikes may also cause injury. 
For example, ball lightning is a rare (and still unexplained) phenomena that may occur as a consequence of lightning strikes and cause injury \cite{ritenour2008lightning}. 
Lightning explosive barotrauma from the accompanying thunder may physically impact a victim and has been known to either push the victim backwards or cause ear damage \cite{ritenour2008lightning, blumenthal2021injuries}.\\

In response to the various possible mechanisms for lightning induced trauma one is advised to adopt the lightning crouch position when outdoors in a thunderstorm.
One should avoid being too close to a tree to avoid side flashes \cite{gookin2010backcountry}, and the potential for pieces of the tree to be launched outwards by the explosive vaporization of sap \cite{christophides2017cardiac}.
Crouching down low may decrease the chance of being directly hit by 50\%; the rolling sphere method (which imagines a sphere whose radius is the striking distance) has been applied to a person standing versus crouching, and a crouching person may also be less likely to initiate a positive upward leader \cite{roeder2008analysis}.
Keeping ones feet close together can also reduce the potential difference across the legs and reduce the risk of current flowing through the body due to this stepped potential \cite{gookin2010backcountry}.
However, for many people squatting continuously with ones feet together may be difficult. 
As such, most people do not place their feet together when performing the lightning crouch position \cite{roeder2014backcountry} and will place their feet apart to help maintain balance \cite{roeder2008analysis}.
The question remains as to whether the difficulty of crouching is worth the effort, or whether the advice should be for people to simply stand up straight and place their feet together \cite{roeder2014backcountry}.
To answer this question we turn to computer simulations.
Given the unpredictable nature of lightning strikes computer models have played a significant part in predicting lightning safety.\\

While some models capture the entire cloud, including the electrification inside clouds and initiation of lightning \cite{mansell2005charge}, for estimating the risk of lightning strikes hitting a given location it is usually considered necessary to just model a smaller spatial region.
Most computer models of lightning have been developed from dielectric breakdown models \cite{niemeyer1984fractal} that have been capable of mimicking the fractal structure of lightning patterns \cite{tsonis1987fractal, mansell2002simulated,guo2021three}.
It is important to note, however, that such models do not include microscopic processes of breakdown \cite{mansell2002simulated}.
That said, the inclusion of randomness and channel tortuosity was considered an improvement over electrogeometric models (such as the Rolling Sphere Method) \cite{petrov2003quantification, riousset2007three}.
Typically, for computational efficiency, a small region of interest is identified and Dirichlet boundary conditions are set at the upper and lower (ground) edges, with Neumann boundary conditions often set for the lateral edges \cite{ioannidis2020fractal, datsios2021stochastic, bian2023quantitative}. Dirichlet boundary conditions are also set at the channel, with the potential either set to that of the cloud above or linearly varying with height due to the finite resistance of the channel \cite{mansell2002simulated, jiang2020simulation, guo2021three}.
The simulation domain is discretized in either two- or three-dimensions, and the electric potential is computed at regular lattice points throughout the domain by solving Laplace's equation using the finite difference method \cite{riousset2007three, ioannidis2020fractal, jiang2020simulation, datsios2021stochastic, guo2021three, bian2023quantitative}.
The boundary conditions vary as the channel propagates down, and nodes in the lattice are assigned a constant electric potential associated with the channel.
The discrete Laplace's equation is then typically solved using the successive over-relaxation method \cite{mansell2002simulated, perera2013fractal}.
To ensure the correct fractal dimension of the lightning channel a probability of lightning progression is typically adopted of the form 
\begin{equation*}
	P_{ij} = \begin{cases}
		P_0  \left(\dfrac{E_{ij}  - E_{cr}}{E_{cr}}\right)^{\eta}  & E_{ij} \geq E_{crit} \\
		0 & E_{ij} < E_{crit}
	\end{cases}
\end{equation*}
where $P_{ij}$ is the probability of the link between spatial nodes $i$ and $j$ becoming part of the channel, $P_0$ is a normalization parameter (lightning is assumed to occur somewhere in the system), $E_{ij}$ is the electric field (the difference in electric potential between nodes $i$ and $j$ divided by the lattice spacing), $E_{cr}$ is the critical electric field below which lightning is assumed to not propagate, and $\eta$ controls the fractal dimension of the lightning pattern \cite{ioannidis2020fractal, datsios2021stochastic, guo2021three}.
Note the critical electric field can be considered height dependent (usually only in systems with large vertical distances), and can be set to different values for downward negative leaders and upward positive leaders \cite{dul1999modeling, datsios2021stochastic}.
A different criteria can also be established for the inception of an upward leader \cite{ioannidis2020fractal, jiang2020simulation}.
The simulation progresses through the iterative addition of segments on to the channel until the lightning channel strikes the ground (or grounded object) \cite{mansell2002simulated}, and through multiple simulations the probability of lightning striking areas of interest can be determined \cite{ioannidis2020fractal, bian2023quantitative}.
It is important to note that the predictive capability of such models is limited as the only experimental validation is in ensuring the fractal dimension of the patterns produced in the simulations is similar to that found experimentally \cite{guo2022validation}.  
Furthermore, the regular grid employed in these models means their is a computational efficiency competition between ensuring a smaller resolution and ensuring a larger simulated domain size \cite{tan2014influence}. 
Bickel \emph{et al.} used an adaptive discretization that reduced the lattice size (by increasing factors of 2) near the areas of interest \cite{bickel2006adaptive}.
The lattice remains regular, however. which might bias lightning propagation directions.
Here, we'll consider an irregular grid that allows the lattice spacing to be varied substantially and removes the regular orthogonal directions of channel progression.
That said, it is worth noting that not all simulations of lightning are limited by a regular lattice structure.\\

The electric potential can be found at any point in space using the superposition of the electric potential as a consequence of charges in the clouds, charges along the lightning channels, and
image charges \cite{gulyas20093d, iudin2017advanced, syssoevmodeling}.
The propagation can then be in any arbitrary direction and to any place in space ahead of a channel \cite{rahiminejad2018fractal, xie2018three}.
However, the addition of charge in close proximity to the point in space often assumes the charge is evenly spaced out over a given region and knowledge of the charge distribution along the channel is required \cite{iudin2017advanced}.
The channel's charge distribution has been assigned different functional forms including a linear charge distribution and a corona tip charge \cite{ait2005lightning} and other functional forms with the charge per unit length at the bottom end of the leader channel being significantly larger than in the rest of the channel \cite{cooray2007lightning, rahiminejad2018fractal}.
Alternatively, the dynamics of charge flow in the channel has been modeled \cite{dul1999modeling, iudin2017advanced}. 
The evolution of the channel conductivity (due to the production and dissipation of Joule heat in the lightning channel) can be coupled with Ohm's law \cite{iudin2017advanced, iudin2018physics, syssoevmodeling}.
The continuity equation for charge density is then used to obtain the charge density along the lightning channel \cite{iudin2018physics}.
It is worth noting that the need for charge density in these superposition models for calculating the electric potential is no different than traditional models that need the electric potential in the channel (typically assuming a linear drop in potential) to calculate the potential using Laplace's equation in the surrounding space.\\

This work contributes to the literature by first developing an irregular 2D lattice model of lightning that involves solving Laplace's equation and lightning propagation on an irregular lattice. The relative risk of being struck by lightning when crouching as opposed to standing is considered. There is a belief that when lightning is attracted by a tall object (such as a tree) then objects nearby might also be more likely to be struck. This is also investigated in the current work.

\section*{Model}

The system size is 2 km by 2 km and discretized with an irregular lattice of points. The lightning channel is assumed to be well established and extend to within a given distance of the ground at the beginning of the simulation. In areas away from where the lightning is likely to strike the density of nodes can be much lower than the areas of interest to minimize the computational expense. The electric potential can be solved in this region of space, and a criteria for extending the lightning channel (based on this electric potential) can be used to capture the extension of the lightning channel. Figure 1 depicts the grid refinement around and within an area of interest. In particular, a lightning strike that extends to 80 m above the ground is initially created. A tree, and a cow standing beneath the tree, are included and as these are the areas of interest the density of nodes increases around these objects and inside these regions to potentially capture the propagation of lightning as it strikes (and potentially passes through) these objects. To my knowledge, all previous models of lightning involving the solution of Poisson's equation have used regular spatial grids and this is the first time an irregular lattice with areas of large density variations has been utilized. \\

Poisson's equation can be solved, given the appropriate Dirichlet boundary conditions (electric potential of the ground and lightning channel), using a Delaunay triangulation of the irregular grids \cite{sukumar2003numerical}. In particular, Poisson's equation becomes 

\begin{equation*}
\nabla \cdot \epsilon \nabla \cdot \phi = \frac{1}{A_i} \left[ \sum_j^n \epsilon_{ij} \alpha_{ij} \phi_j - \left(\sum_j^n \epsilon_{ij} \alpha_{ij} \right) \phi_i \right] = 0
\end{equation*}
where $\epsilon$ is the permittivity, $\phi$ is the electric potential, $A_i$ is the area of a Voronoi cell, $\epsilon_{ij}$ is the permittivity associated with the link between nodes $i$ and $j$, $\phi_i$ is the electric potential at node $i$, and $\alpha_{ij}$ is the following Laplace weight
\begin{equation*}
 \alpha_{ij} = \frac{s_{ij}}{l_{ij}}
\end{equation*}

where $s_{ij}$ is the length of the Voronoi edge associated with, and $l_{ij}$ is the distance between, nodes $i$ and $j$. These distances are depicted in Fig. 1d. A system of 250000 nodes is solved using the successive over-relaxation scheme (commonly used for solving Poisson's equation) until the relative change in potential summed over the system is less than $1\times10^{-5}$. \\

The permittivity between nodes $i$ and $j$ can be determined from the permittivities at the nodes
\begin{equation*}
 \frac{2}{\epsilon_{ij}} = \frac{1}{\epsilon_i} + \frac{1}{\epsilon_j}
\end{equation*}
where $\epsilon_i$ is the permittivity associated with node $i$. This allows us to increase the permittivity in objects from that of the surrounding air. For example, this would lower the potential in a tree and make the lightning channel mode likely to strike the tree than the surrounding ground. This is equivalent to including space charge on the surface of the object \cite{feynman2010feynman}. It is worth noting that the relative permittivity of a tree will depend heavily on moisture content \cite{torgovnikov1993dielectric} and can be higher for skin and muscle tissue \cite{michel2015accuracy}; here a value of 10 is taken for the tree and 100 is taken for the person.\\

Lightning can propagate in this system in a number of ways; the channel can extend downwards, the channel might branch and create another path, or the inception of an upwards lightning channel can occur from the ground (or objects on the ground) \cite{datsios2021stochastic}. The probabilities are taken to be similar to that of the dielectric breakdown model of lightning \cite{niemeyer1984fractal, pietronero1988physical}. To ensure a more isotropic propagation of the lightning channel, the component of the electric field in the direction of a lattice connection is used to calculate the probability of a channel being established in this direction. 
\begin{equation*}
	P_{ij} \propto \begin{cases}
		P_0 P_{\beta} \left(\dfrac{(\vec{E}_i \cdot \hat{r}_{ij})  - E_{crit}}{E_{crit}}\right)^{\eta}  & (\vec{E}_i \cdot \hat{r}_{ij}) \geq E_{crit} \\
		0 & (\vec{E}_i \cdot \hat{r}_{ij}) < E_{crit}
	\end{cases}	
\end{equation*}
where $P$ is the probability of lightning progression between nodes $i$ and $j$, $P_0$ is a constant that depends on whether the lightning is traveling through air or another material, $P_{\beta}$ is a branching factor that depends on the lattice density, $E_{i}$ is the electric field, $E_{crit}$ is the critical electric field which may depend on height (air density) and the material the lightning is propagating in, and $\eta$ is the exponent that controls the locality of lightning progression; larger values of $\eta$ and $E_{crit}$ favor a single channel with little branching. $\hat{r}_{ij}$ is a unit vector between nodes $i$ and $j$. The electric field is obtained from the Taylor expansion of the electric potential 
\begin{equation*}
\phi_j - \phi_i \approx \vec{E}_i \cdot \vec{r}_{ij}
\end{equation*}
There are as many equations (Taylor expansions) associated with a node as there are connections in the lattice. The electric field can be calculated at each node as the least squares solution to this overdetermined system of equations. Obtaining the electric field from all neighboring nodes in this way has been found to produce lightning channels whose propagation does not appear to be influenced by variations in lattice density. To separately vary the probability of branching we also introduce the branching factor, given by
\begin{equation*}
	P_{\beta} = \begin{cases}
		\beta r_{ij} & \text{if extension adds to the branching}\\
		1 & \text{if extension extends the channel}
	\end{cases}	
\end{equation*}
where $\beta$ is a constant (zero would suppress all branching, which has been considered previously for upward leaders) and $r_{ij}$ is the length of channel potentially being added to node $i$. Again, this ensures the predicted lightning channel is independent of the variations in density in the underlying lattice (otherwise more dense regions might exhibit more branching). Only sites that are currently part of the channel have a probability of extending the channel. \\

In addition to the extension of (or branching off from) existing channels, upward leaders may be initiated from the ground or other objects. The probability of upward leader inception is given by
\begin{equation*}
	P_{ij} \propto \begin{cases}
		P_0 \left(\dfrac{E_{ij} - E_{incept}}{E_{incept}}\right)^{\eta}   & E_{ij} \geq E_{incept} \\
		0 & E_{ij} < E_{incept}
	\end{cases}	
\end{equation*}
where $P_0$ is a constant, and $E_{incept}$ is the critical electric field required for upward leader inception. The critical upward leader inception electric field may depend on surface roughness and the objects curvature at length scales smaller than the lattice spacing. \\

A bond is randomly selected given the above probabilities. In particular, the probabilities are normalized to ensure the sum of probabilities is 1 and the cumulative distribution is used to select the next event that will occur in the system. The i\textsuperscript{th} event will occur when the cumulative distribution satisfies $c_i < \text{RND}[0,1] < c_{i+1}$, where $c_i$ is the cumulative distribution (sum of probabilities up to event $i$), $\text{RND}[0,1]$ is a random number between 0 and 1, and the chance of an event going over this random number is proportional to the size of the probability being added to the cumulative distribution. \\

As new channels are added the electric potential must be assigned as part of the Dirichlet boundary conditions for Poisson's equation. In particular, if a new site is created at node $j$ from node $i$ then the new electric potential is given by
\begin{equation*}
	\phi_j = \phi_i + E_{ch} l_{ij}
\end{equation*}
where $\phi_i$ is the electric potential at the existing site, $E_{ch}$ is the electric field along the channel, and $r_{ij}$ is the distance between sites $i$ and $j$. For upward leaders the electric field along the channel is not included and $\phi_j = \phi_i$. \\

At each iteration, therefore, a new site is chosen given the above probability to be added to the channel. The electric potential is then assigned to this site, assuming a constant electric field in the lightning channel. Given the new site and the potential (now a part of the Dirichlet boundary conditions) we solve Poisson's equation and calculate the electric potential for all points on the lattice. The electric field is then determined from the gradients in electric potential and new probabilities of sites being added to the channel are predicted based on the local electric fields. 
Given the probabilities a new site is chosen to be added to the channel and the simulation iterates through these steps until the lightning has intercepted the ground, an object or an upward leader.

\section*{Results}

In the current simulations, the potential at the top of the simulation is set to \qty{40e6}{\V} and the potential at the ground is zero. $E_{crit} = \qty{50e3}{\V\per\m}$, $E_{incept} = \qty{500e3}{\V\per\m}$, and $E_{ch} = \qty{0}{\V\per\m}$. A ``notch'' is used to initiate the simulation, consisting of a channel that extends to within \qty{80}{\m} of the ground. The parameters $\eta$ and $\beta$ control the fractal dimension of the lightning channels and are varied to mimic real lightning channel geometries.\\

Figure 2 shows both the fractal dimension (Fig. 2a) and the length of the channels (Fig. 2b) as a function of varying both $\eta$ and $\beta$. For each set of parameters the fractal dimension and length are averaged over 100 simulations. Increasing $\eta$ has been associated with decreasing the fractal dimension. 
Here, with $\beta = 1$, the fractal dimension for $\eta = 1$ is 1.47 and this decreases to 1.31, 1.23, and 1.19 as $\eta$ increases to 2, 3, and 4, respectively. In other words, a higher exponent causes probabilities to increases more with increasing electric field, and more strongly favors the extension of a single channel in an area of high electric field. A lower exponent, in contrast, would result in relatively similar probabilities in multiple areas of similar electric field and the simultaneous growth of multiple channels and branches would be expected. In the current model we also vary the branching factor, $\beta$. If a possible extension to the lightning would simply extend a channel then the probability is unchanged. However, if the extension is in a new direction (creating a new branch) then the probability may be increased or decreased by a factor proportional to $\beta$. For example, in Fig. 2a increasing $\beta$ from 0.125 to 4 (while keeping $\eta = 1$) resulted in a fractal dimension that increased from 1.31 to 1.53. The fractal dimension, therefore, is not just dependent on one parameter in the current model but two parameters. Fig. 2b depicts the length of the lightning channels (including all branches and not just the main channel that intersects with the ground) and elucidates the effects of varying both $\eta$ and $\beta$. The effects of increasing $\eta$ can be seen to dramatically reduce the length of the lightning channel. In effects the lightning exhibits less branching and a single straight channel might emerge. Increasing $\beta$ would obvious increase the amount of branching, and result in much larger lengths of lightning channel in the system. The fractal dimension and length and not entirely correlated, however.\\

As an example, Figure 3 depicts lightning patterns from three simulations. The fractal dimensions of all three systems are similar; 1.314, 1.319, and 1.318, respectively. However, the length of the lightning channels decreases with values of $745.4\,\text{m}$, $683.7\,\text{m}$, and $538.8\,\text{m}$, respectively. The difference between the parameters in these systems is that the value of $\eta$ increases (1, 2, and 3, respectively), and the value of $\beta$ increased also (0.125, 1, and 4, respectively). In other words, the system in Fig. 3a favors the extension of multiple channels (including branching) with low $\eta$ while the low $\beta$ was decreasing the probability of branching. In Fig. 3b the system favors the extension of the main channel due to the high $\eta$ while also promoting branching (but not necessarily the extension of these branches) with high $\beta$. The parametrization of similar models often involves mapping the fractal dimension of the simulated lightning to that of natural lightning (typically assumed to be in the range of 1.1 to 1.4 \cite{kawasaki2000does, miranda2016multifractal, guo2022validation}); in the following simulations we take $\eta = 3$ and $\beta = 4$, although this choice is arbitrary.\\

The model is now applied to predicting the probability of a direct strike hitting either a standing person, a crouching person, or a cow. Fig. 4a shows the cumulative distribution function for the probability of lightning striking a given location. In other words, if there was a large number of strikes in a given location then this would present in the cumulative distribution function as a region on the graph with a larger slope. Given the standing person, crouching person and cow where all placed in the center of the system (\qty{80}{\m} below the lightning channel) there does not appear to be any indication that the person (whether standing or crouching) or cow would be any more likely to be hit by the lightning than the flat ground. This can also be seen from Figs. 4b to 4d. The person (or cow) and ground are colored according to the number of strikes that hit the location during the simulations. The strike pattern appears random with no discernible effects from the person (standing or crouching) or the cow.
More simulations (this data was obtained from 200 independent simulations with unique lattice structures) might reveal a small difference in the probability of the lightning hitting the slightly elevated figures, but from the simulations here and given the parameters in the model there does not appear to be any statistically significant likelihood of any of these figures being hit. It is also possible that the statistical significance might present itself if the lightning was taken to be closer than \qty{80}{\m} above the ground when the simulation was initiated, or that the probability of upward leader inception was too low in the current model. \\

Figure 5 depicts the cumulative distribution function and strike pattern for three systems consisting of a tree and a person standing either \qty{10}{\m}, \qty{15}{\m}, or \qty{20}{\m} from the tree. There is a perception that standing near an object that might attract lightning will increase ones chance of being struck. From the cumulative distribution function we can see that the slope of the graph does increase in the center (where the tree is located). In particular, given the increase in slope, the tree is 3.1 times more likely to be hit than the same locations in the absence of the
tree (the simulation of lightning strikes hitting the flat land). However, there is no indication that the person in close proximity is any more or less likely of being hit by lightning that other areas. Figs. 5b to 5d show the strike patterns, with a significant number of strikes (darker coloring) occurring on the tree than away from the tree. It is worth noting that the distribution of lightning strikes is still quite noisy and more simulations might result in smoother data. However, given the current model there does not appear to be an increased risk of being hit by lightning via a direct strike when standing near the tree. 

\section*{Conclusions}

A new model of lightning that captures lightning progression on an irregular lattice is presented. The model has been used to capture the probability of direct lightning strikes on a person standing, a person crouching, and a cow in an open plain. Furthermore, the probability of a person being hit directly when near a tree was also elucidated. Direct lightning strikes, however, only contribute to a relatively small number of lightning injuries or deaths and future models will capture the probabilities of other modes of lightning induced trauma. Another limitation of the current model is that it is limited to two-dimensions. While the extension of the model to three-dimensions would be trivial, it is perhaps unnecessary. Models that use the charge distributions (in the clouds, the ground and the lightning channel itself) to calculate the electric field ahead of the progressing lightning channel do not rely on a lattice (regular or irregular) \cite{gulyas20093d, iudin2017advanced, syssoevmodeling}. 
Rather than imagine what the electric potential might be in the lightning channel and use this as the boundary condition to Poisson's equation, such models imagine what the charge distribution might be and use superposition to estimate the potential or electric field in areas of interest; the benefit being that it is only necessary to solve in areas of interest and it removes the computational expense of solving Poisson's equation. 
Such models have incorporated charge flow to estimate the charge distribution, which allows the models to also acquire a timescale \cite{iudin2017advanced}. 
However, the probability of the lightning progression can also be thought of as a rate, which would allow all of these models of lightning to be assigned a timescale; towards the end of the simulation, when electric fields intensify ahead of the propagating lightning channel, one would then expect the timescale to be much smaller and the lightning would travel faster. 
While the speed of the lightning does increase slightly as the lightning channel approaches the ground \cite{rakov2007lightning}, this does not appear to be significant enough to assign drastically increased rates of lightning progression.  The lightning probability presumably depends on more than just the electric field. 

Currently the physics of lightning progression is captured by simply matching the fractal dimension of the simulations (by varying the exponent, $\eta$) to the wide range of values seen in a limited number of experimental studies. This has resulted in a wide range of values adopted for $\eta$ \cite{perera2013fractal, xie2018three, rahiminejad2018fractal, ioannidis2020fractal, datsios2021stochastic, guo2021three}, and as we have seen here the fractal dimension can also be varied by including an additional parameter (in this case the branching factor, $\beta$).
Recent observations of streamers and space stems ahead of the propagating lightning channels indicates a much more complicated (and probably time-dependent) process than is considered in these simple models \cite{qi2016high, rakov2016fundamentals, khounate2021insights, saba2022close}. Future work will incorporate the evolution of streamer zones ahead of the lightning channels and the formation of space stems ahead of the leader tips.

\clearpage
\section*{Figures}

\includegraphics[width=\linewidth]{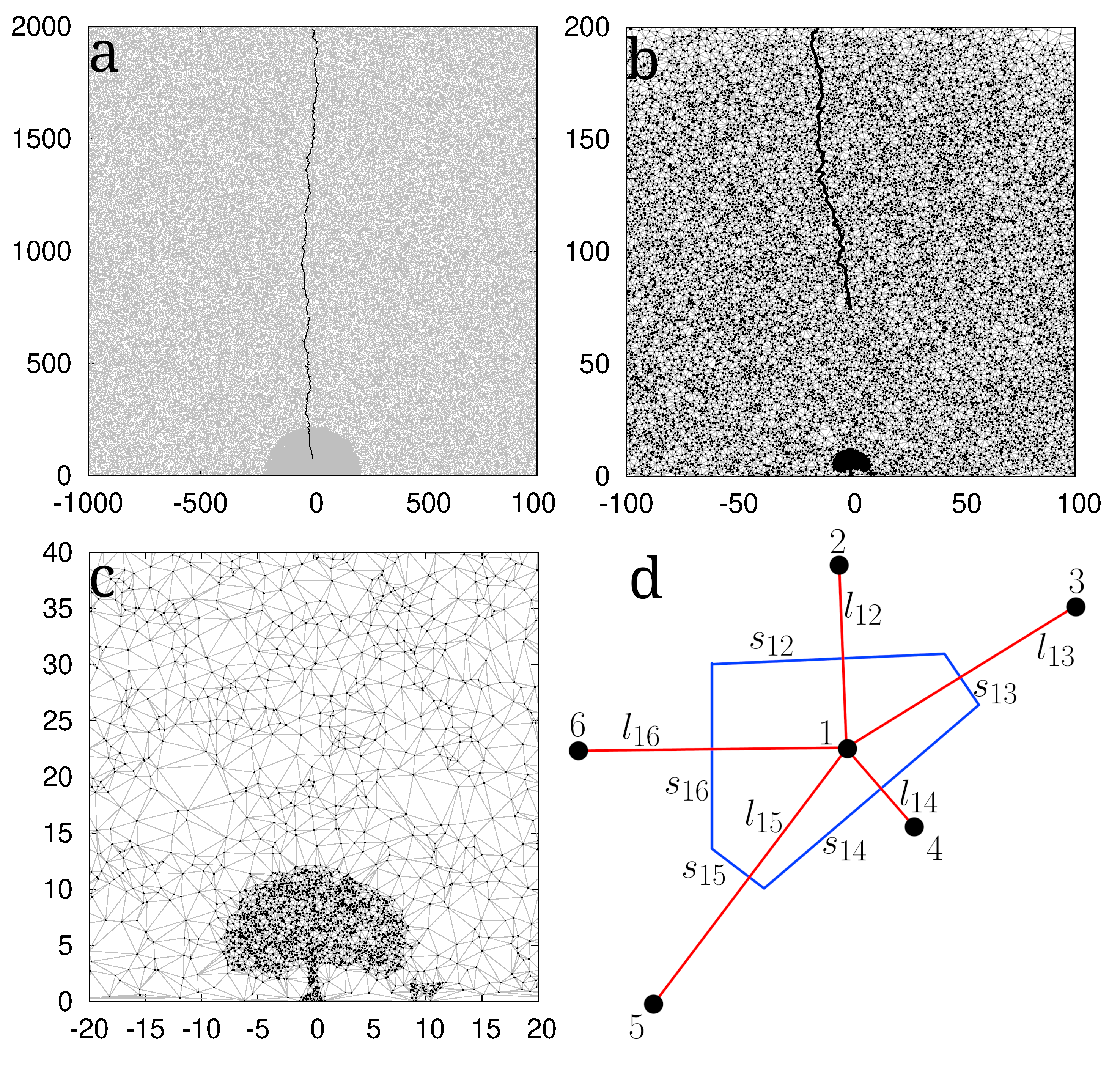} 
Figure 1: The lattice refinement around areas of interest. a) The lightning channel is initially assumed to extend to 80 m above the ground. The density of lattice points is much lower away from the area of interest. b) The density of lattice points increases near the area of interest. c) Within the objects being captured the refinement of nodes can be increased further. d) The distances between nodes, $l_{ij}$, and the length of the edges of a Voronoi cell, $s_{ij}$, used to calculate the Laplace weights for solving Poisson's equation on this irregular grid.
\clearpage

\begin{center}
\includegraphics[width=0.7\linewidth]{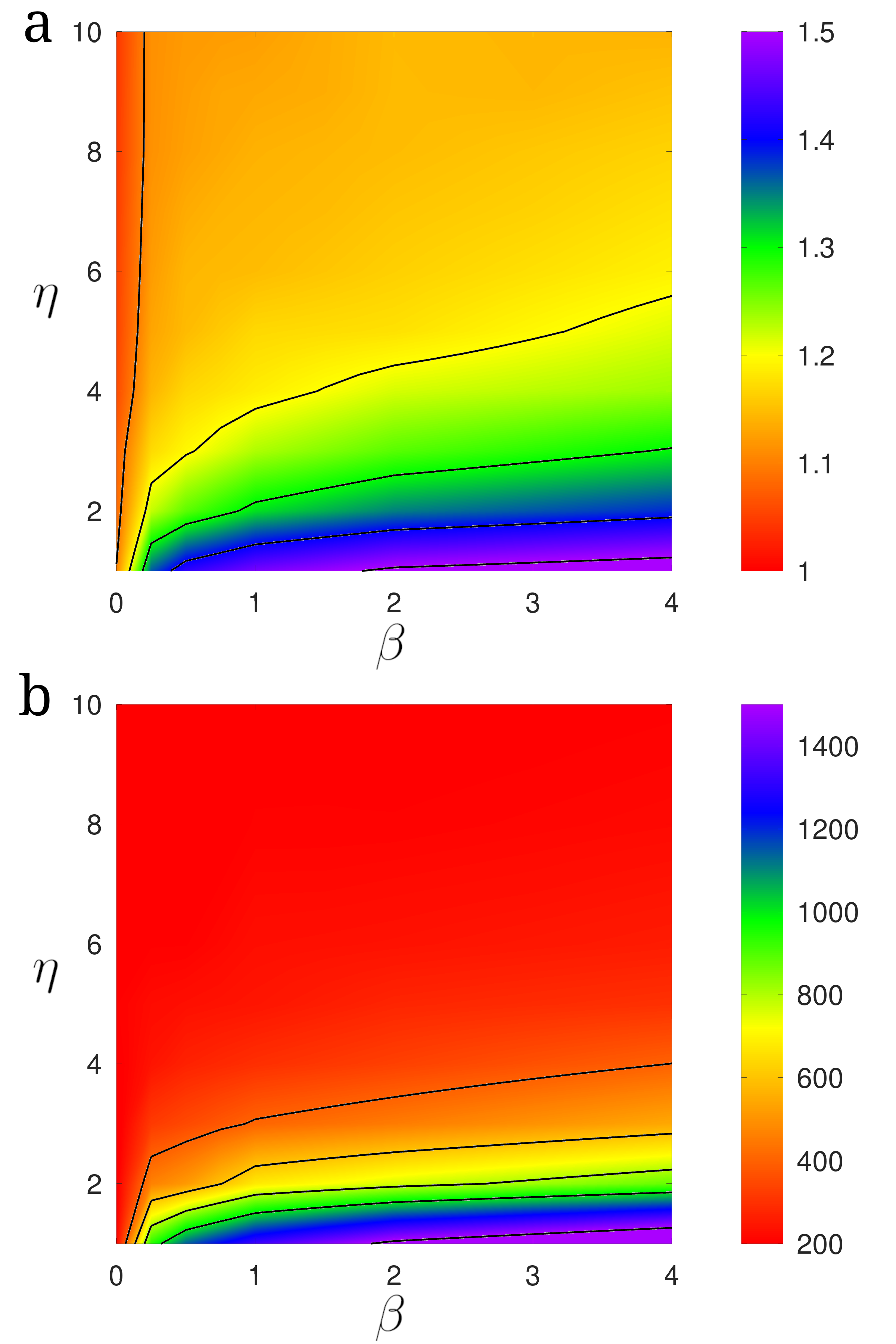} 
\end{center}
Figure 2: The effects of varying the probability of branching, $\beta$, and the exponent, $\eta$, on a) the fractal dimension of the lightning patterns generated, and b) the length of the lightning channels. For each set of parameters the fractal dimension and the length are averaged over 100 simulations using different lattice networks.
\clearpage

\begin{center}
\includegraphics[width=0.65\linewidth]{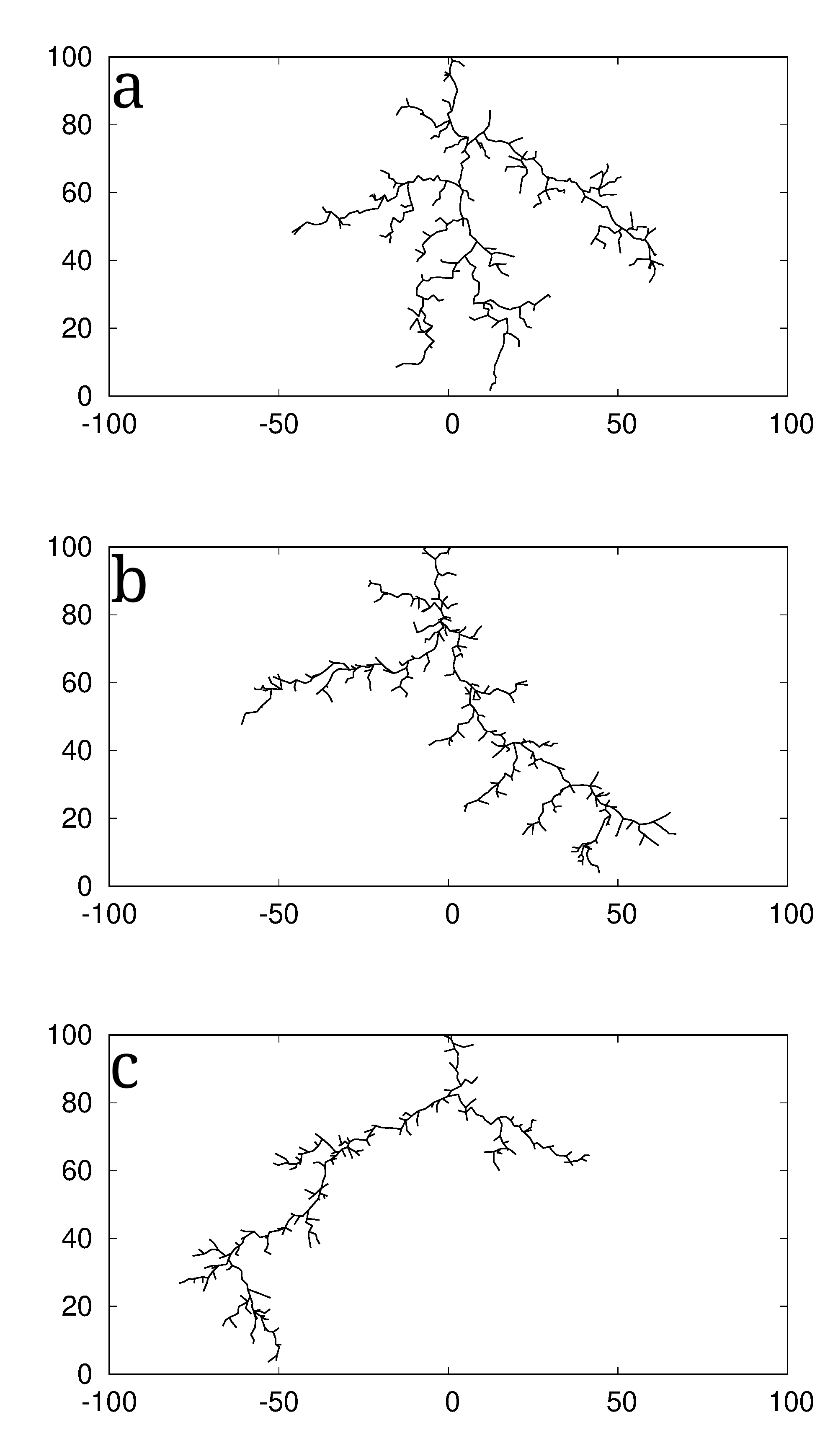} 
\end{center}
Figure 3: Lightning patterns are depicted from three simulations. a) $\beta = 0.125$ and $\eta = 1$. The fractal dimension is 1.314 and length is $745.4\,\text{m}$. b) $\beta = 1$ and $\eta = 2$. The fractal dimension is 1.319 and length is $683.7\,\text{m}$. c) $\beta =4$ and $\eta = 3$. The fractal dimension is 1.318 and length is $538.8\,\text{m}$. For these parameters the average fractal dimensions where also similar; $1.307 \pm 0.036$, $1.312 \pm 0.032$, and $1.303 \pm 0.035$, respectively.
\clearpage

\begin{center}
\includegraphics[width=0.65\linewidth]{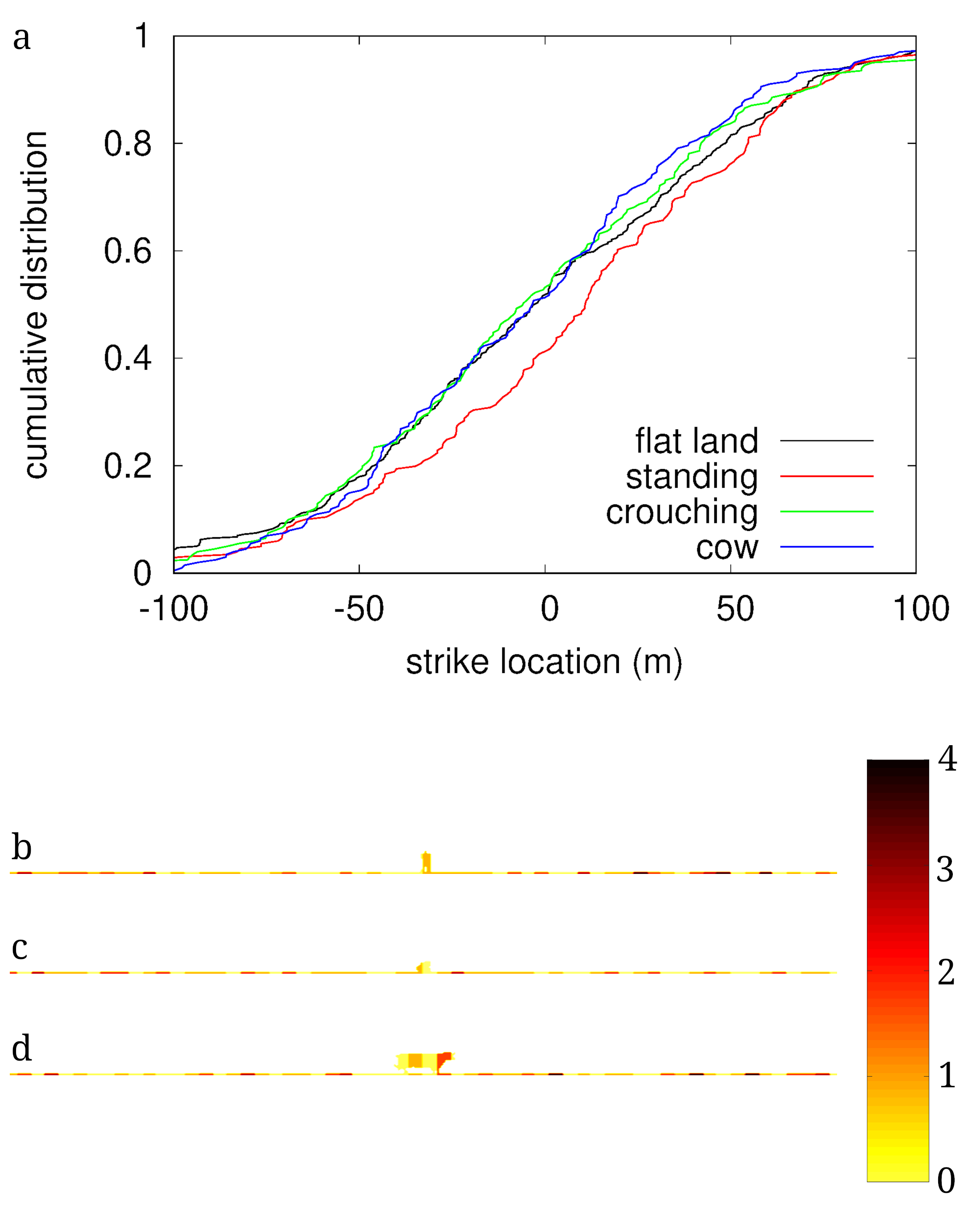} 
\end{center}
Figure 4: The probability of smaller objects being directly hit by a lightning channel that was assumed to have already propagated to within 80 m. a) The cumulative distribution of strike location for a flat topography, a person standing, a person crouching, and a cow. The number of hits (out of 200 runs) at different locations are also exhibited for b) a person standing, c) a person crouching, and d) a cow.
\clearpage

\begin{center}
\includegraphics[width=0.65\linewidth]{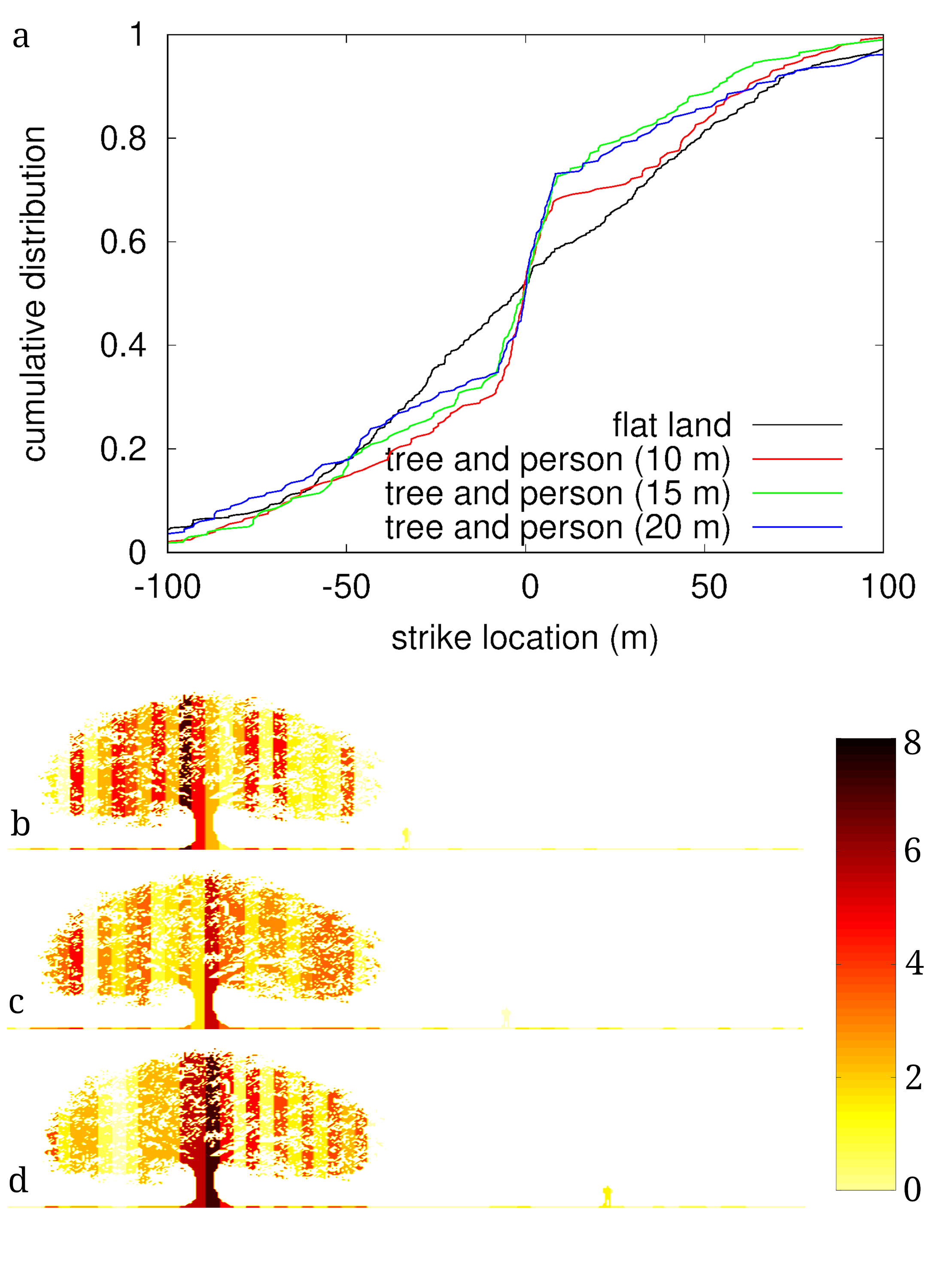} 
\end{center}
Figure 5: The probability of a person being directly hit by lightning that was assumed to have already propagated to within 80 m of a nearby tree. a) The cumulative distribution of strike location for a flat topography, and system with a tree and a person standing varying distances from the tree. The number of hits (out of 200 runs) at different locations are also exhibited for b) a tree with a person standing 10 m away, c) a tree with a person standing 15 m away, and d) a tree with a person standing 20 m away.
\clearpage

\bibliographystyle{plain}
\bibliography{buxton.bib}

\end{document}